\documentstyle{article}
\begin{document}
\setlength{\baselineskip}{15pt}
\title{Algebraic recasting of nonlinear systems of ODEs into universal 
   formats}
\author{ Benito Hern\'{a}ndez--Bermejo$^{\; a}$  \and V\'{\i}ctor 
         Fair\'{e}n$^{\; a ,\; *}$ \and L\'{e}on Brenig$^{\; b}$}
\date{}

\maketitle

{\em \mbox{} \\
$^{a}$ Departamento de F\'{\i}sica Fundamental. Universidad Nacional de 
Educaci\'{o}n a Distancia. Senda del Rey S/N, 28040 Madrid, Spain. \\
$^{b}$ Service de Physique Statistique, Plasmas et Optique non Lineaire. 
Universit\'{e} Libre de Bruxelles. CP 231, Boulevard du Triomphe, 1050 
Brussels, Belgium.}

\mbox{}

\begin{abstract}
   It is sometimes desirable to produce for a nonlinear system of ODEs a new 
representation of simpler structural form, but it is well known that this 
goal may imply an increase in the dimension of the system. This is what 
happens if in this new representation the vector field has a lower degree of 
nonlinearity or a smaller number of nonlinear contributions. Until now both 
issues have been treated separately, rather unsystematically and, in some 
cases, at the expense of an excessive increase in the number of dimensions. 
We unify here the treatment of both issues in a common algebraic framework. 
This allows us to proceed algorithmically in terms of simple matrix 
operations. 
\end{abstract}

\mbox{}


% 02.30H = Ordinary differential equations.
% 03.65F = Algebraic methods (in 03 = Classical and Quantum Physics).

\mbox{}

Short title: Algebraic recasting.

\mbox{}

\mbox{}

\mbox{}

\mbox{}

$^{*}$ Author to whom all correspondence should be addressed. Email \\ 
vfairen@uned.es

\pagebreak

\newtheorem{mitma}{Theorem}
\newtheorem{lm}{Lemma}
\newtheorem{co}{Corollary}

\begin{flushleft}
{\bf 1. Introduction}
\end{flushleft}

Very often we may be interested in transforming a given system of nonlinear 
ODEs into some specified format.
Some of these formats are, given their very simple structure, 
particularly attractive. One is the generalized Riccati format ---or the 
Lotka-Volterra equations, as one of its variants. There are many other 
useful formats: The Generalized 
Mass Action, of use in modelling chemical reactions or in metabolic control 
theory \cite{vsi}; S-systems, devised to capture the saturable 
characteristics of nonlinear systems \cite{vsi}; the unimonomial canonical 
form \cite{bro}, which has proved to be suitable for analyzing integrability 
properties; etc. 

Reformatting a system of ODEs finds its more rewarding 
field of application in the obtainment of approximate solutions, especially 
numerical. It is in the field of Celestial Mechanics where this approach has 
been first systematically carried out \cite{els}, by rewritting equations of 
motion into polynomial form and expanding solutions in terms of a 
Taylor-like integration scheme. This procedure has been shown to 
lead to time-saving and highly accurate integrators \cite{flc,abz}, 
specifically adapted to symbolic manipulation \cite{rcp}. Irvine {\em et 
al.\/} \cite{dhi} have also argued in favour of reformatting before 
proceeding to a numerical integration. Actually, additional constants of 
motion introduced in some recasting procedures may prove useful for 
exploiting certain families of numerical algorithms used in symplectic or 
Poisson formulations \cite{qis}: By preserving exactly first integrals we 
may indirectly measure the accuracy of the numerical approximation. 

Loosely speaking, two will essentially be the ultimate goals of reformatting 
procedures: i) an equivalent system of lower degree  of nonlinearity 
---quadratic at most, for example; ii) or alternatively, a vector field with 
a smaller number of contributions ---terms--- in its components. Generally, 
the system must be embedded into some higher-dimensional set of equations in 
order to be recasted into the target format. This happens if  
the manipulation aims at a reduction in the degree of the 
nonlinearity defining the vector field. Examples are: The Carleman embedding 
\cite{car,kys}, where an $n$-dimensional system of nonlinear ODEs is set into 
the form of an infinite-dimensional linear system, or Kerner's procedure 
\cite{ker} for reducing nonlinearities to quadratic form. 

Resort to an embedding may also be the case if we wish instead to reduce the 
number of nonlinear terms building up the components of the vector field. 
This is necessary in the study of certain integrability properties of ODEs 
in parameter space \cite{bro}. Also, in some modelling approaches in theoretical 
biochemistry \cite{vo1,bv2}, where net rates of variation in concentration of 
species are split into the sum of a single inflow and a single outflow terms. 
Finally, in the case of half-systems and binary half-systems \cite{vo1}, 
rates are reduced to a single contribution per equation. 

The problem with the different procedures used nowadays in recasting is that 
they are rather unsystematic, relying too much on a rule of thumb approach to 
the problem. In this article we wish to treat the issue from a systematic 
point of view. We choose as an appropriate framework to do so a format for 
the vector field defined in terms of quasipolynomial expressions, which has 
proven to be suitable for representing general nonlinear vector fields 
\cite{bv2,byv}. It has been indifferently called Generalized Lotka-Volterra 
format \cite{bro}, Generalized Mass Action systems \cite{vo1}, multinomial 
differential systems \cite{pym}, power-law rates systems \cite{bv3} and 
polynomial differential systems \cite{gou}. The distinctive property of the 
quasipolynomial format is that it is form-invariant under certain classes 
of transformations, called quasimonomial and new-time transformations. 
Consequently, its analysis can be carried out in terms of simple matrix 
algebra. 

The structure of the article is as follows. We outline in Subsections 2.1 and 
2.2 the basic algebraic properties of systems of quasipolynomial ODEs, which 
will define our operational framework. In Subsections 2.3 to 2.5 we 
demonstrate how an arbitrary quasipolynomial system can be reduced to a 
standard form which will constitute the starting point in later sections. 
Sections 3 and 4 are the core of this work: We explain and 
illustrate the reduction procedure for the degree of nonlinearity and number 
of nonlinear terms, respectively. We give some concluding remarks in 
Section 5. 

\begin{flushleft}
{\bf 2. The quasipolynomial formalism}
\end{flushleft}

We shall consider a quasipolynomial (QP) system of ODE's of the form 
\cite{bre}
\begin{equation}
   \dot{x}_i = x_{i}(\lambda _{i} + \sum_{j=1}^{m}A_{ij}\prod_{k=
      1}^{n}x_{k}^{B_{jk}}) , \;\:\;\: i = 1 \ldots n 
   \label{eq:glv}
\end{equation}
where $n$ and $m$ are positive integers, and $A$, $B$ and $\lambda$ are 
$n \times m$, $m \times n$ and $n \times 1$ real matrices, respectively. 
Sometimes it will be also convenient to consider the $n \times (m+1)$ 
composed matrix $M = ( \lambda \mid A)$. We shall also assume 
that the $x_i$ are real and positive (if this is not the case, an appropriate 
phase-space translation is to be performed on the variables in order to 
ensure positiveness, see \cite{byv}). In what follows, $n$ will always denote 
the number of variables of a QP system, and $m$ the number of the 
quasimonomials
\begin{equation}
   \prod_{k=1}^{n}x_{k}^{B_{jk}} , \;\:\;\: j = 1 \ldots m .
\end{equation}

\begin{flushleft}
{\em 2.1 Dimension preserving transformations}
\end{flushleft}

System (\ref{eq:glv}) is formally invariant under quasimonomial 
transformations \cite{bro}
\begin{equation}
   x_{i} = \prod_{k=1}^{n} y _{k}^{C_{ik}} , \;\: i=1,\ldots ,n
   \label{bec}
\end{equation}
for any invertible real matrix $C$. Under (\ref{bec}), matrices $B, A, 
\lambda$ and $M$ change to 
\begin{equation}
\label{tqm4}
B' = B \cdot C \;\:, \:\;\: A' = C^{-1} \cdot A \;\: , \:\;\: 
\lambda ' = C^{-1} \cdot \lambda \;\: , \:\;\: M' = C^{-1} \cdot M \;\: ,
\end{equation}
respectively, but the QP format is preserved. 
An $(n,m)$-family of QP systems in thus split into classes of 
equivalence, the product $B \cdot M$ being a class invariant. 
Under the previous assumptions, the solutions of 
all the members of the class are topologically equivalent \cite{byv}. 

The quasimonomial transformations are complemented by new-time 
transformations of the form \cite{bre2,gor2}
\begin{equation}
    \label{ntt}
    dt \; = \; \left( \prod_{i=1}^n x_i^{\beta _i} \right) \; dt' \;\;\: , 
\end{equation}
where the $\beta _i$ are real constants. Equation (\ref{ntt}) also 
preserves the QP format. 

\begin{flushleft}
{\em 2.2 Transformations that do not preserve dimension} 
\end{flushleft}

A QP system can be subjected to some interesting additional manipulations 
that do not preserve the actual dimension. 

We may on the one hand embed a system (\ref{eq:glv}) into a larger QP system 
by means of an expansion of matrices $B$ and $M$ up to any arbitrary $p$
\begin{equation}
     B \longrightarrow \tilde{B} 
               = ( B \mid B_{m \times p} ) \;\; , \: \; \: \; 
     M \longrightarrow \tilde{M}
               = \left( \begin{array}{c} 
               M \\ O_{p \times (m+1)} 
           \end{array} \right) \;\; ,
     \label{v10}
\end{equation}
with arbitrary entries for matrix $B_{m \times p}$. Here, $O_{p \times (m+1)}$ 
is the $p \times (m+1)$ null matrix. By this embedding we add $p$ new 
variables which remain constant under the action of the expanded vector field 
(\ref{v10}), for all $t \geq 0$. The original equations are not altered if 
the initial values for these $p$ additional variables are chosen to be equal 
to one.

On the other hand, another transformation can be defined as
\begin{equation}
     B \longrightarrow \tilde{B} 
               = ( B \mid O_{m \times p} ) \;\; , \: \; \: \; 
     M \longrightarrow \tilde{M}
               = \left( \begin{array}{c} 
               M \\ M_{p \times (m+1)} 
           \end{array} \right) 
     \label{v9}
\end{equation}
The entries of $M_{p \times (m+1)}$ are in principle arbitrary. By this 
operation we simply add to the original system (\ref{eq:glv}) $p$ variables 
in such a way that in the expanded system the original $n$ variables are not 
coupled to these $p$ new variables. 

Without loss of generality, we shall assume in forthcoming sections that 
the QP system under study has $m \geq n$, and that the 
ranks of matrices $A$ and $B$ are both maximal (i.e. rank($A$) $=$ rank($B$) 
$= \; n$). In what remains of the section we proceed to demonstrate for the 
first time that any QP system not fulfilling these requirements can be 
reduced to this standard situation. 

The algorithm proceeds in three steps:

\noindent {\bf 1)} Reduction to the case $m \geq n$, by means of a 
quasimonomial transformation which reduces the solution for some of the 
variables to a quadrature.

\noindent {\bf 2)} When $m \geq n$, reduction to a system with maximal rank of 
$B$. The method is similar to that in (1).

\noindent {\bf 3)} When $m \geq n$ and rank($B$) is maximal, reduction to a 
system with maximal rank of $A$. There are two steps:

{\bf 3.1)} Reduction to maximal rank of $M$ by means of a quasimonomial 
transformation, followed by the decoupling of some variables. 

{\bf 3.2)} Reduction to maximal rank of $A$ by means of a new-time 
transformation. 

\begin{flushleft}
{\em 2.3 Reduction to the case $m \geq n$}
\end{flushleft}

Assume a QP system for which $m < n$. Let rank($B$) $=r$, with $r \leq m$. 
Since $B$ is an $m \times n$ matrix, dim $\{$ ker($B$) $\}$ $= \; (n-r) > 0$. 
Consequently, we can perform a quasimonomial transformation of matrix
\begin{equation}
   \label{mmn}
   C = \left( \begin{array}{c} 
            I_r \\ \mbox{} \\ O_{(n-r) \times r}
       \end{array} \right| 
       \left. \begin{array}{ccc}
            \mbox{}        & \mbox{} & \mbox{}     \\
            \phi ^{(r+1)}  & \ldots  & \phi ^{(n)} \\
            \mbox{}        & \mbox{} & \mbox{} 
       \end{array} \right)
\end{equation}
where $I_r$ is the $r \times r$ identity matrix, and $\{ \phi ^{(r+1)}, 
\ldots , \phi ^{(n)} \}$ is a set of column vectors which constitute a basis 
of ker($B$). When applied, the quasimonomial transformation given by matrix 
(\ref{mmn}) leads to $B' =$ $( B'_{m \times r} \mid O_{m \times (n-r)})$. 
This implies that we are led to an 
$r$-dimensional QP system for $x_1 , \ldots , x_r$, plus ($n-r$) quadratures 
for $x_{r+1} , \ldots , x_n.$ The problem is thus reduced to the case 
$m \geq n$ (with $n=r$). 

\begin{flushleft}
{\em 2.4 Reduction to a system with maximal rank of B}
\end{flushleft}

Let us assume now a QP system with $m \geq n$, and rank($B$) not maximal. 
As Brenig and Goriely have shown \cite{bre2}, the system can be transformed 
into an equivalent one for which rank($B$) is maximal. The technique is 
similar to that in Subsection 2.3: If $r \;=$ rank($B$), $r<n$, then 
($n-r$) variables can be decoupled by means of transformation (\ref{mmn}). 
The result now is a QP system of $r$ variables and $m$ quasimonomials plus 
($n-r$) quadratures. 

\begin{flushleft}
{\em 2.5 Reduction to a system with maximal rank of A}
\end{flushleft}

Here we shall proceed in two steps. First, we shall 
demonstrate how a QP system can be reduced to an equivalent one for which 
rank($M$) is maximal. Then, the scope of the result shall be enlarged to 
the reduction to a system with maximal rank of $A$. As before, we assume 
$m \geq n$.

We solve the first part of the problem by giving two new constructive 
theorems:

\begin{mitma}
\label{thy}
{\rm If rank($M$) $= r < n$ in equations (\ref{eq:glv}), then there exist 
$(n-r)$ time-independent first integrals of the system. Moreover, these first 
integrals are functionally independent, and thus the system evolves on a 
manifold of dimension $r$.}
\end{mitma}

\begin{flushleft}
{\em Proof.}
\end{flushleft}

We can assume, without loss of generality, that the $r$ first rows of $M$ are 
the linearly independent ones. Then, there exist real constants $\gamma _{ki}$, 
$1 \leq i \leq r$, $r+1 \leq k \leq n$, such that:
\begin{equation}
   M_{kl} = \sum_{i=1}^{r} \gamma_{ki} M_{il} \;, \;\: 1 \leq l \leq m+1  \; .
\end{equation}
From the form of equations (\ref{eq:glv}), this implies:
\begin{equation}
   \frac{\dot x_k}{x_k} = \sum_{i=1}^{r} \gamma_{ki} \frac{\dot x_i}{x_i} \;.
\end{equation}
After a simple integration this leads to the set of $(n-r)$ first 
integrals:
\begin{equation}
   x_k^{-1} \prod_{i=1}^{r} x_i^{\gamma _{ki}} = c_k \; , 
   \label{clp}
\end{equation}
where the $c_k $ are real constants given by the initial conditions. Their 
functional independence is ensured if the rank of the corresponding Jacobian 
$F$ is maximal. It is straightforward to see that as rank($F$) $= (n-r)$, 
the $(n-r)$ first integrals are functionally independent.$\Box$

We can now state:

\begin{mitma}
\label{cglv}
   {\rm Consider the $n$-dimensional QP system (\ref{eq:glv}) of matrix 
   $M$. If \linebreak rank($M$) $= r < n$, then there exists a quasimonomial 
   transformation that leads from such a system to an $n$-dimensional QP 
   system of matrix 
   \begin{equation}
      \label{mcc}
      M_r = \left( \begin{array}{c}
               R_{r \times (m+1)} \\
               O_{(n-r) \times (m+1)}
            \end{array} \right) 
   \end{equation} }
\end{mitma}
Since the proof relies on simple matrix algebra, we shall omit it.

It should be emphasized that, if rank($B$) is maximal for the original 
system, this will be also the case for the target QP system of $r$ variables 
and $m$ quasimonomials. Consequently, the reduction to maximal rank of $M$  
has been completed. 

It is still possible that rank($M$) is maximal, but rank($A$) is not ---for 
$m \geq n$, rank($M$) $= \; n$ and rank($A$) $= \; (n-1)$. A new-time 
transformation of the form (\ref{ntt}) will solve it if, in particular, we 
choose $\beta _i = -B_{ji}$, $i= 1, \ldots ,n$, for an appropriate value of 
$j$, $1 \leq j \leq m$. The result is a QP system for which both matrices 
$M$ and $A$ have maximal rank $n$. Notice that this new-time transformation 
does not change the rank of matrix $B$. 

Consequently, we have demonstrated that every QP system can be reduced to 
an equivalent QP system with $m \geq n$ and matrices $B$, $A$ and $M$ of 
maximal rank $n$. From now on, it is assumed that the system under 
consideration complies to this standard form.

\pagebreak
\begin{flushleft}
{\bf 3. Reduction of the degree of nonlinearity}
\end{flushleft}

\begin{flushleft}
{\em 3.1 Algebraic theory}
\end{flushleft}

In this subsection we shall show how the degree of nonlinearity can be 
reduced by means of appropriate manipulations of a given QP system, the 
ultimate goal being a quadratic vector field. 

The simplest case arises when $m=n$: $B$ is an $n \times n$ 
invertible matrix, and we can set $C = B^{-1}$. The result is another flow 
for which $B' = I_n$, namely a Lotka-Volterra system. The reduction is thus 
achieved.

In the complementary case $m > n$, $B$ is not square and cannot be reduced 
to diagonal form. The maximal reduction of the degree of nonlinearity 
proceeds as follows: We split $B$ in two submatrices:
\begin{equation}
    B = \left( \begin{array}{c}
                  B_{n \times n} \\ B_{(m-n) \times n}
               \end{array} \right)
\end{equation}
Since rank($B$) $=n$ we can assume, without loss of generality, that 
rank($B_{n \times n}$) $=n$. Then, if $C = B_{n \times n}^{-1}$, the 
result is a QP system for which:
\begin{equation}
    B' = B \cdot C = 
               \left( \begin{array}{c}
                  I_n \\ B_{(m-n) \times n} \cdot B_{n \times n}^{-1}
               \end{array} \right)
\end{equation}
In other words, only $n$ quasimonomials in the vector field can be reduced to 
quadratic form in this case. The form of the $(m-n)$ remaining quasimonomials 
is fixed by $B_{(m-n) \times n} \cdot B_{n \times n}^{-1}$ and a reduction of 
the degree cannot be applied to them. 

When $m>n$ a complete reduction of the original system to a Lotka-Volterra 
form is accomplished by performing an embedding: The process leading to 
(\ref{v10}) particularized to $p = m-n$. Here, $B_{m \times (m-n)}$ in 
(\ref{v10}) must be composed of arbitrary entries such that $\tilde{B}$ is 
invertible. Then we have defined an expanded $m$-dimensional QP system 
and an 
invertible matrix $\tilde{B}$ to which the procedure of the $m = n$ case can 
be applied. Consequently, a unique Lotka-Volterra representative can also be 
associated to system (\ref{eq:glv}) in the case $m>n$: That of matrix 
$M_{LV} = \tilde{B} \cdot \tilde{M} = B \cdot M$.

We can now state the following new result for these systems:

\begin{mitma}
\label{n8j}
{\rm The target Lotka-Volterra system has $(m-n)$ independent first 
integrals, given  by  }
\begin{equation}
    F_i (\xi _1 , \ldots , \xi _m) = \xi_i^{-1}\ \prod_{j=1}^{n} 
    \xi_j^{\alpha_{ij}} -1=0 \;, \;\: n+1 \leq i \leq m
    \label{vlv}
\end{equation}
{\rm where the constants $\alpha _{ij}$, $n+1 \leq i \leq m$, $1 \leq 
j \leq n$, satisfy the relations:  }
\begin{equation}
   B_{il} = \sum_{j=1}^{n} \alpha_{ij} B_{jl} \;, \;\: 1 \leq l \leq n
\end{equation}
\end{mitma}
The proof is not especially involved, and we omit it.

Embedding (\ref{v10}) can actually be applied in such a way that the number 
$k$ of new variables is smaller than $(m-n)$. 
Therefore, it will be possible to reduce to quadratic nonlinearity at most 
$(n+k)$ quasimonomials by means of a quasimonomial transformation. 
On the other hand, the resulting $(n+k)$-dimensional QP system 
will have $k$ independent first integrals of quasimonomial form 
introduced by the embedding. Rank($M$) $=n$ for the target system 
and the explicit characterization of these $k$ first integrals is given 
by Theorem \ref{thy}. 

To summarize, the alternative steps are:

\noindent {\bf a)} If the dimension is to be preserved: One quasimonomial 
transformation.

\noindent {\bf b)} In those cases ($m>n$) where the dimension may be 
increased: Add extra variables of constant value 1 and make an appropriate 
quasimonomial transformation.

\begin{flushleft}
{\em 3.2 Example 1}
\end{flushleft}

We shall illustrate the previous procedures with a system of 
relevance in the field of molecular physics, the Morse oscillator \cite{byj}.
The system is given by:
\begin{eqnarray}
\dot{x} & = & y  \nonumber \\
\dot{y} & = & -2d\alpha e^{-\alpha x} (1 - e^{-\alpha x})
\end{eqnarray}
We now perform a phase-space translation of magnitude $c$, followed by the 
introduction of a new variable $z = e^{- \alpha x}$. Thus: 
\begin{eqnarray}
\dot{x} & = & x[x^{-1}y-cx^{-1}]  \nonumber \\
\dot{y} & = & y[ay^{-1}z-aby^{-1}z^2]  \label{21spm} \\
\dot{z} & = & z[ \alpha c - \alpha y] \nonumber
\end{eqnarray}
where $a = -2db\alpha $ and $b = e^{\alpha c}$.

The maximal possible reduction in degree without embedding corresponds to the 
choice:
\begin{equation}
   C = \left( \begin{array}{ccc}
              -1 & 1 & 0 \\ -1 & 0 & 0 \\ 0 & -1 & 1
              \end{array} \right) ^{-1} 
\end{equation}
After the transformation, the result is another QP system with exponents:
\begin{equation}   
   B' = \left( \begin{array}{ccc}
        1 & 0 & 0 \\ 0 & 1 & 0 \\ 0 & 0 & 1 \\ 1 & -1 & 2 \\ 1 & -1 & 0 
               \end{array} \right) 
\end{equation}
We see that the last two quasimonomials have not been simplified by the 
transformation, as was to be expected.

We consider now the reduction via the Lotka-Volterra embedding. First, notice 
that from matrix $B$ of (\ref{21spm}), if we call $\xi _i$ the 
quasimonomials, with $i = 1 , \ldots , 5$,  we have 
\begin{equation}
\label{xibh}
\xi _4 = \xi _1 \xi_2^{-1}\xi_3^2 \;\: , \;\:\;\: \xi _5 = \xi _1 \xi _2^{-1} 
\end{equation}
Since the quasimonomials of the original QP system are the variables of the 
Lotka-Volterra system, these constraints will be present as first integrals 
in the final set of equations. We expand the characteristic matrices of 
(\ref{21spm}) according to (\ref{v10}). After this embedding, the 
Lotka-Volterra system is given by matrix:
\begin{equation}
   M_{LV} =  
       \left( \begin{array}{cccccc} 
        0        & -1 & c &  a & -ab & 0         \\
        0        & -1 & c &  0 &  0  & 0         \\
        \alpha c &  0 & 0 & -a &  ab & - \alpha  \\ 
      2 \alpha c &  0 & 0 & -a &  ab & -2 \alpha \\
        0        &  0 & 0 &  a & -ab & 0
       \end{array} \right) 
\end{equation}
We observe that rank($M_{LV}$) $=3$. In particular, we have (row4) $=$ (row1) 
$-$ (row2) $+$ 2$\times$(row3) and (row5) $=$ (row1) $-$ (row2). After a 
straightforward integration of the corresponding evolution equations (see 
Theorem \ref{thy} for further details) we recover the firts integrals 
(\ref{xibh}). This description of the system has the advantage of being 
purely quadratic. 

We can also proceed from (\ref{21spm}) to a partial embedding, for example 
the one for which the expanded matrices are:
\begin{equation}
   \tilde{M} = \left( \begin{array}{cccccc}
       0    & 1  & -c  & 0  &  0  &  0       \\
       0    & 0  &  0  & a  & -ab &  0       \\
   \alpha c & 0  &  0  & 0  &  0  &  -\alpha \\
       0    & 0  &  0  & 0  &  0  &  0 
              \end{array}   \right)  \;\: , \;\:\;\:
   \tilde{B} = \left( \begin{array}{cccc}
         -1    & 1  & 0 & 0 \\
         -1    & 0  & 0 & 0 \\
          0    & -1 & 1 & 0 \\
          0    & -1 & 2 & 1 \\
          0    & 1  & 0 & 0 
       \end{array} \right) 
\end{equation}
plus the initial condition $x_4 (0) = 1$. We now take
\begin{equation}
   C = \left( \begin{array}{cccc}
              -1 & 1 & 0 & 0 \\ -1 & 0 & 0 & 0 \\ 0 & -1 & 1 & 0 \\ 0 & -1 
              & 2 & 1 
              \end{array} \right) ^{-1} 
\end{equation}
This leads to the following system:
\begin{eqnarray}   
\dot{z}_1 & = & z_1[ -z_1 +cz_2 +\alpha z_3 -abz_4]  \nonumber \\
\dot{z}_2 & = & z_2[ -z_1 +cz_2 ]  \\
\dot{z}_3 & = & z_3[\alpha c -\alpha z_3 +ab z_4 -\alpha z_1z_2^{-1}]   \nonumber \\
\dot{z}_4 & = & z_4[2\alpha c -\alpha z_3 +ab z_4 -2\alpha z_1z_2^{-1}] \nonumber 
\end{eqnarray}
where four quasimonomials have been reduced. From matrix $C$ we have 
$x_4 = x_4 (0) = 1 = z_1^{-1} z_2 z_3^{-2} z_4$, a first integral which is 
present in the target system, as inferred from the fact that (row4) $=$ 
(row1) $-$ (row2) $+$ 2$\times$(row3). In this case we have a balance 
between the increase in the dimension and the expression of the flow in 
quadratic terms. 

\begin{flushleft}
{\em 3.3 Example 2}
\end{flushleft}

As a second example, we shall consider a model describing sustained 
time-oscillations in the concentration of electron-hole pairs ($x_1$) and 
excitons ($x_2$) in an intrinsic semiconductor \cite{cyv}. The process 
consists of the following steps:
\begin{eqnarray}
   \mbox{Photogeneration of carriers:} & & \gamma 
      \stackrel{g}{\longrightarrow} e+h  \\
   \mbox{Stimulated production of excitons:} & & e+h+x_2 
      \stackrel{c}{\longrightarrow} 2x_2 \\
   \mbox{Radiative decay of excitons:} & & x_2 \stackrel{k}{\longrightarrow} 
      \gamma 
\end{eqnarray}
When high-order kinetics is allowed in the model, the process is described 
by the equations:
\begin{eqnarray}
\dot{x}_1 & = & g - cx_1^2x_2  \nonumber \\
\dot{x}_2 & = & cx_1^2x_2 - \frac{kx_2}{(1+qx_2)^m} \label{msc1}
\end{eqnarray}
System (\ref{msc1}) can be readily put into QP form by introducing a new 
variable such as $x_3=(1+qx_2)^{-1}$. The resulting flow is:
\begin{eqnarray}
\dot{x}_1 & = & x_1 [ gx_1^{-1} - cx_1x_2 ] \nonumber \\
\dot{x}_2 & = & x_2 [ cx_1^2 - kx_3^m ]  \label{msc2} \\
\dot{x}_3 & = & x_3 [ -cqx_1^2x_2x_3 + qkx_2x_3^{m+1} ] \nonumber 
\end{eqnarray}
The three interactions originally present in (\ref{msc1}) correspond to the 
second, third and fourth quasimonomials of the QP system (\ref{msc2}). 
More precisely, these quasimonomials can be seen as interactions {\em per 
capita,\/} as we have already pointed out \cite{bv3}. Accordingly, we can 
particularize such interactions as the representative variables, by means of 
a quasimonomial transformation of matrix:
\begin{equation}
   C = \left( \begin{array}{ccc}
              1 & 1 & 0 \\ 2 & 0 & 0 \\ 0 & 0 & m 
              \end{array} \right) ^{-1} 
\end{equation}
The transformed equations are:
\begin{eqnarray}
\dot{y}_1 & = & y_1 [ -cy_1 + cy_2 -ky_3+gy_2^{-1/2}] \nonumber \\
\dot{y}_2 & = & y_2 [ -2cy_1 +2gy_2^{-1/2} ]       \label{msc3} \\
\dot{y}_3 & = & y_3 [ -cqmy_1y_2^{1/2}y_3^{1/m} + 
                       qkmy_1y_2^{-1/2}y_3^{1+1/m}] \nonumber 
\end{eqnarray}
In this representation, the three {\em per capita\/} interactions (given by 
$y_1$, $y_2$ and $y_3$) appear as linear quasimonomials. This means that we 
can alternativetely describe the system in terms of rates of stimulated 
production and radiative decay of excitons, i.e., the fundamental physical 
processes, and this can be done while reducing the degree of nonlinearity of 
the model.

We can complete the reduction to quadratic (Lotka-Volterra) form. The 
resulting matrix is:
\begin{equation}
   \label{lve2s}
   A_{LV} =  
       \left( \begin{array}{cccccc} 
        -g &   c & 0 &  0 &  0       & 0       \\
         g &  -c & c & -k &  0       & 0       \\
        2g & -2c & 0 &  0 &  0       & 0       \\ 
         0 &   0 & 0 &  0 & -cqm     & qkm     \\
        2g & -2c & c & -k & -cq      & qk      \\
         0 &   0 & c & -k & -cq(1+m) & qk(1+m)
       \end{array} \right) 
\end{equation}
From matrix (\ref{lve2s}), the first integrals are 
$y_2y_3^{1/2}y_4^{1/m}y_5^{-1}$, $y_2y_3^{-1/2}y_4^{1+1/m}y_6^{-1}$ and 
$y_1y_3^{1/2}$. This conclusion can also be anticipated from matrix $B$ of 
(\ref{msc2}), by following the same procedure than in Example 1. 

The last possibility is that of a partial embedding. We may, for instance, 
add one new variable:
\begin{equation}
   \tilde{M} = \left( \begin{array}{ccccccc}
      0 & g  & -c  &  0  & 0  &  0  & 0    \\
      0 & 0  &  0  &  c  & -k &  0  & 0    \\
      0 & 0  &  0  &  0  & 0  & -cq & qk   \\
      0 & 0  &  0  &  0  & 0  &  0  & 0  
              \end{array}   \right) , \;
   \tilde{B} = \left( \begin{array}{cccc}
            -1 & 0 & 0   & 1 \\
             1 & 1 & 0   & 0 \\
             2 & 0 & 0   & 0 \\
             0 & 0 & m   & 0 \\
             2 & 1 & 1   & 0 \\
             0 & 1 & m+1 & 0 
       \end{array} \right) 
\end{equation}
Our quasimonomial transformation is now the one given by:
\begin{equation}
   C = \left( \begin{array}{cccc}
              -1 & 0 & 0 & 1 \\ 1 & 1 & 0 & 0 \\ 
               2 & 0 & 0 & 0 \\ 0 & 0 & m & 0 
              \end{array} \right) ^{-1} 
\end{equation}
The final system is then:
\begin{eqnarray}
\dot{z}_1 & = & z_1 [ -gz_1 + cz_2  ] \nonumber \\
\dot{z}_2 & = & z_2 [ gz_1 - cz_2 + cz_3 - kz_4 ]  \label{mscf} \\
\dot{z}_3 & = & z_3 [ 2gz_1 - 2cz_2 ] \nonumber \\
\dot{z}_4 & = & z_4 [ -cqmz_2z_3^{1/2}z_4^{1/m} + 
                qkmz_2z_3^{-1/2}z_4^{1+1/m} ] \nonumber 
\end{eqnarray}
In this case, we describe the system by looking at the full set of 
interactions, including also the {\em per capita\/} rate of photogeneration 
of carriers ($z_1$). It can be easily checked that the first integral 
$z_1z_3^{1/2}$ has been introduced in the process. However, now the degree 
of nonlinearity of the description is smaller than in (\ref{msc3}).

\begin{flushleft}
{\bf 4. Reduction of the number of nonlinear terms}
\end{flushleft}

\begin{flushleft}
{\em 4.1 Algebraic theory}
\end{flushleft}

The complementary approach to the problem of structural simplification of a 
nonlinear system proceeds to the reduction of the total number of nonlinear 
terms constituting each component of the vector field. 

For $m=n$, it is possible to redistribute the quasimonomials assigning one of 
them to each equation. $A$ is invertible and we can therefore set $C = A$: 
$A' = I_n$ for the target flow, yielding a system for 
which only one nonlinear term per equation is present. This system is called 
{\em unimonomial system}. The reduction to this form has been useful in the 
study of integrability properties of ODEs \cite{bro} as well as in the 
construction of models in theoretical biochemistry \cite{bv2}. 

In the case $m>n$ it is no longer possible to reallocate $m$ quasimonomials 
in $n$ equations by means of a quasimonomial transformation. The maximal 
reduction that can be achieved in this way proceeds as follows: Let $A$ be: 
$A =$ $( A_{n \times n} \mid A_{n \times (m-n)})$. 
Since rank($A$) is maximal, we can assume that $A_{n \times n}$ is 
invertible. Then, if we set $C = A_{n \times n}$ (a choice that determines 
$C$ completely) the result is a QP system with $A' =$ $( I_n \mid A_{n 
\times n}^{-1} \cdot A_{n \times (m-n)})$. 

In other words, only $n$ quasimonomials can be doled out among the $n$ 
equations in this case. The redistribution of the remaining $(m-n)$ 
quasimonomials cannot be controlled and depends on the entries of the initial 
matrix $A$.

It is necessary to increase to $m$ the number of equations, before a complete 
redistribution of the $m$ quasimonomials, by means of transformation 
(\ref{v9}), particularized to $p=(m-n)$. The entries of $M_{(m-n) \times (m+1)}$ 
in (\ref{v9}) are in principle arbitrary, but should nevertheless be chosen 
appropriately to ensure maximal rank for the expanded matrix $\tilde{A}$. 
The problem is thus reduced to the previous situation $m=n$. 

The price for this simplification of the system is twofold. 
First, the dimension has been increased. Second, the information from the 
original system is intermingled with spurious components originated from the 
embedding process, in such a way that both expanded and original systems are 
not topologically equivalent: It is clear from equation (\ref{v9}) that the 
information is not necessarily confined to a submanifold of the 
$m$-dimensional phase space, as was the case in the LV embedding. 
Consequently, the original topology of the starting $n$-dimensional phase 
space will not be preserved, in general, in an $n$-dimensional submanifold of 
the target $m$-dimensional phase space; the matter being further worsened 
after any quasimonomial transformation acting upon the expanded system. 

We shall analyze in detail this problem here for the first time. 
We shall demonstrate that, although 
the topological equivalence is not preserved in the process, the information 
corresponding to the original system can always be retrieved in each step of 
the procedure by means of a simple projection technique. 

Let us make some definitions; we will call ${\bf x}$ the $m$-dimensional 
vector 
\begin{equation}
{\bf x} = (x_1 (t),\ldots , x_n (t), \overbrace{1, \ldots ,1}^{m-n} 
\; )^T \;\: ,
\end{equation}
which is essentially the solution to the original $n$-dimensional problem 
(\ref{eq:glv}), and 
\begin{equation}
{\bf x}_{exp} = (x_1 (t),\ldots , x_m (t))^T \;\: ,
\end{equation}
the corresponding solution of the expanded system (\ref{v9}) (particularized 
to $p=m-n$). Since all variables are assumed to be strictly 
positive, we can equivalently deal with the logarithms $\ln {\bf x}$ and 
$\ln {\bf x}_{exp}$ of the vectors. Then we can retrieve the solution 
${\bf x}$ from ${\bf x}_{exp}$ by means of the matrix identity 
\begin{equation}
   \ln {\bf x} = P_n \cdot \ln {\bf x}_{exp} \;\: ,
   \label{pn}
\end{equation}
where $P_n =$ diag($I_n$, $O_{(m-n) \times (m-n)}$). $P_n$ just projects, in 
logarithmic space, the $m$-dimensional solution onto the hyperplane 
$\ln x_{n+1} = \ldots = \ln x_{m} = 0$. 

We can now consider the effect of a quasimonomial transformation performed 
over system (\ref{v9}) with $p=m-n$. We will denote as ${\bf z}, \; 
{\bf z}_{exp}$, the transformed trajectories from ${\bf x}, \; {\bf x}_{exp}$, 
respectively. The following relations hold: 
\begin{equation}
  \ln {\bf x} = C \cdot \ln {\bf z} \; , \;\:\; 
  \ln {\bf x}_{exp} = C \cdot \ln {\bf z}_{exp} 
\end{equation}
From this, together with (\ref{pn}), the following identity can be deduced: 
\begin{equation}
  \ln {\bf z} = C^{-1} \cdot P_n \cdot C \cdot \ln {\bf z}_{exp} \equiv 
  P_n' \cdot \ln {\bf z}_{exp} \;\: .
  \label{pc}
\end{equation}
Equation (\ref{pc}) shows how $P_n$ is transformed under a quasimonomial 
transformation of matrix $C$. It is then clear from definition (\ref{bec}) 
that ${\bf z}$ belongs to the submanifold of ${\cal R}^m$
\begin{equation}
   \prod _{j=1}^m z_j^{C_{ij}} = 1 \; , \;\: i = n+1, \ldots , m \; ,
   \label{stc}
\end{equation}
or 
\begin{equation}
   \sum _{j=1}^m C_{ij} \ln z_j = 0 \; , \;\: i = n+1, \ldots , m \; .
   \label{hp}
\end{equation}
In fact, this submanifold is the one obtained by transforming the 
surface containing the original $n$-dimensional phase space, namely $x_{n+1} 
= \ldots = x_m = 1$, by means of the quasimonomial transformation. It is not 
difficult to see that $P_n'$, as given by equation (\ref{pc}), is nothing 
else than the projection matrix on hyperplane (\ref{hp}) of ${\cal R}^m$: 
Substituting (\ref{pc}) into (\ref{hp}) we have 
\begin{equation}
   \sum _{j=1}^m C_{ij} \ln ({\bf z}_n)_j = \sum _{j=1}^m (P_n \cdot C)_{ij} 
   \ln ({\bf z}_m)_j = 0 \; \; \mbox{ if } \; i>n \; ,
\end{equation}
since $(P_n \cdot C)_{ij} = 0$ if $i>n$. Consequently, this demonstrates that 
the original phase space is topologically equivalent to the {\em projection \/}  
of the transformed trajectories $\ln {\bf z}_m (t)$ onto the $n$-dimensional 
manifold (\ref{hp}). However, it must be pointed out that this does {\em not\/} 
imply that the single quasimonomial form embedding preserves topological 
equivalence between the original $n$-dimensional phase space and some 
particular submanifold of ${\cal R}^m$. Nevertheless, this constructive 
procedure permits to discern relevant from spurious information at every 
stage of the manipulation of the equations. In particular, no inversion of 
the embedding procedure is required for the retrieval of the significant 
information of the system: The explicit knowledge of the projection matrix 
$P_n'$ allows direct access to it, once the solution ${\bf z}_m (t)$ has been 
integrated. 

We end this subsection by adding that the unimonomial embedding can be 
performed only partially. Let $k$ be the number of extra equations in 
(\ref{v9}), $1 \leq k < (m-n)$. In this case, only $(n+k)$ 
quasimonomials will be amenable of redistribution among the equations. It is 
also clear that appropriate projection operators must be introduced in order 
to screen the relevant information.

Again, we summarize the main alternative steps:

\noindent {\bf a)} If the dimension is to be preserved: A quasimonomial 
transformation. 

\noindent {\bf b)} If the dimension can be increased, the sequence is: 

{\bf b.1)} Add new variables decoupled from the original ones. 

{\bf b.2)} Make an appropriate quasimonomial transformation. 

{\bf b.3)} ``Keep track'' of the relevant information: Find the projection 
matrix. 

\begin{flushleft}
{\em 4.2 Example 3}
\end{flushleft}
   
Let us consider the Brusselator equations \cite{nyp}, a model system 
for the Belousov--Zhabotinskii reaction. In QP form, the equations are:
\begin{eqnarray}
   \dot{x}_1 & = & x_1[-(b+1) + x_1x_2 + ax_1^{-1}]  \nonumber \\
   \dot{x}_2 & = & x_2 [bx_1x_2^{-1} + x_1^2]        \label{55spm}
\end{eqnarray}

If we proceed without embedding, two quasimonomials can be redistributed at 
will among the equations. We may select the quasimonomial transformation of 
matrix:
\begin{equation}
   C = \left( \begin{array}{cc}
              1 & 0 \\ 0 & b 
              \end{array} \right) 
\end{equation}
For the reduced QP system we have: 
\begin{equation}
\label{57spm}
   M' = \left( \begin{array}{ccccc} 
                -(b+1) & 1 & 0 & a &   0     \\ 
                    0  & 0 & 1 & 0 & - 1/b 
              \end{array} \right) 
\end{equation}

On the contrary, if we want a complete redistribution of the quasimonomials, 
an embedding must be applied, for instance the following one: 
\begin{equation}
   \label{ex1}
   \tilde{M} = \left( \begin{array}{ccccc} 
               - (b+1) & 1 & 0 & a &  0  \\ 
                   0   & 0 & b & 0 & -1  \\
                   0   & 0 & 0 & 1 &  0  \\
                   0   & 0 & 0 & 0 &  1  
              \end{array} \right) \;\:\; , \;\;\:
   \tilde{B} = \left( \begin{array}{cccc}
              1 &  1 & 0 & 0 \\ 
              1 & -1 & 0 & 0 \\ 
             -1 &  0 & 0 & 0 \\ 
              2 &  0 & 0 & 0 
              \end{array} \right) 
\end{equation}
The final unimonomial system, which is reached after a quasimonomial 
transformation of matrix $C = \tilde{A}$, is: 
\begin{eqnarray}
   \dot{z}_1 & = & z_1 [ -(b+1) + z_1 z_2^b z_3^a z_4^{-1} ] \nonumber \\
   \dot{z}_2 & = & z_2 [ z_1 z_2^{-b} z_3^a z_4 ]            \nonumber \\
   \dot{z}_3 & = & z_3 [ z_1^{-1} z_3^{ -a} ]                \label{um1} \\
   \dot{z}_4 & = & z_4 [ z_1^2 z_3^{2a} ]                    \nonumber
\end{eqnarray}
This is an optimal flow, in the sense that only one nonlinear term is 
present per equation. We retrieve the original trajectories of 
system (\ref{55spm}) if we project the logarithms 
of the solutions of (\ref{ex1}) onto the plane $\{ \ln (x_3)=0, \; 
\ln (x_4)=0 \}$, i.e., by means of the projection matrix: 
\begin{equation}
  P_2 = \left( \begin{array}{cc}
              I_2 & \mbox{} \\ \mbox{} & O_{2 \times 2}
              \end{array} \right) 
\end{equation}
Under the quasimonomial transformation leading to the reduced system 
(\ref{um1}) this operator is transformed to:
\begin{equation}
  P_2' = C^{-1} \cdot P_2 \cdot C = 
            \left( \begin{array}{cccc}
              1 &  0 &  a &  0   \\ 
              0 &  1 &  0 & -1/b \\ 
              0 &  0 &  0 & 0    \\
              0 &  0 &  0 & 0  
            \end{array} \right) 
\end{equation}
This operator still projects onto the plane $\{ \ln (z_3)=0, 
\; \ln (z_4)=0 \}$. Consequently, although the flow is four-dimensional, only 
the knowledge of the first two variables is necessary for the investigation 
of the topology of the initial flow (\ref{55spm}), since such topology is 
equivalent to that obtained when $z_1(t)$ and $z_2(t)$ are represented on 
${\cal R}^2$. Another way of expressing this is by noticing that the 
projection plane $\{ \ln (x_3)=0, \; \ln (x_4)=0 \}$ remains invariant under 
the quasimonomial transformation we perform. 

The third alternative is that of a partial embedding. One possibility for 
adding the new variable is:
\begin{equation}
   \tilde{M} = \left( \begin{array}{ccccc} 
               - (b+1) & 1 & 0 & a &  0  \\ 
                   0   & 0 & b & 0 & -1  \\
                   0   & 0 & 0 & 1 &  0 
              \end{array} \right) \;\:\; , \;\;\:
   \tilde{B} = \left( \begin{array}{cccc}
              1 &  1 & 0 \\ 
              1 & -1 & 0 \\ 
             -1 &  0 & 0 \\ 
              2 &  0 & 0 
              \end{array} \right) 
\end{equation}
The partially reduced system is then: 
\begin{eqnarray}
   \dot{y}_1 & = & y_1 [ -(b+1) + y_1 y_2^b y_3^a ]          \nonumber \\
   \dot{y}_2 & = & y_2 [ y_1 y_2^{-b}y_3^a - (1/b) y_1^2 y_3^{2a} ]   \label{um2} \\
   \dot{y}_3 & = & y_3 [ y_1^{-1} y_3^{-a} ]                 \nonumber
\end{eqnarray}
This system represents a compromise between (\ref{57spm}) and (\ref{um1}). 
The projection plane $\{ \ln (x_3)=0 \}$ is again invariant under the 
quasimonomial transformation:
\begin{equation}
  P_2 = \left( \begin{array}{ccc}
              1 &  0 & 0 \\ 
              0 &  1 & 0 \\ 
              0 &  0 & 0 
            \end{array} \right) \Longrightarrow 
  P_2' = C^{-1} \cdot P_2 \cdot C = 
            \left( \begin{array}{ccc}
              1 &  0 &  0 \\ 
              0 &  1 & -1 \\ 
              0 &  0 &  0 
            \end{array} \right) 
\end{equation}
Again the third variable is to be discarded.

It should be emphasized that, along this example, a great simplification of 
the procedure given in Subsection 4.1 comes from the fact that the projection 
plane remains invariant under the quasimonomial transformation and, 
consequently, the projection procedure reduces, in the final system, to the 
suppression of the spurious variables. Such feature is {\em not\/} particular 
to this example: The unimonomial embedding can always be chosen in such a 
way that this property holds. 

\begin{flushleft}
{\em 4.2 Example 4}
\end{flushleft}
   
We shall now apply our technique to a system which models the nonlinear 
interaction of three waves (see \cite{bre} and references therein):
\begin{eqnarray}
   \dot{x}_1 & = & x_1 \left[ \lambda _1 + \sum_{j=1}^3 N_{1j} x_j^2 + 
                 \gamma x_1^{-1} x_2 x_3 \right] \nonumber \\
   \dot{x}_2 & = & x_2 \left[ \lambda _2 + \sum_{j=1}^3 N_{2j} x_j^2 \right] \label{s3ob} \\
   \dot{x}_3 & = & x_3 \left[ \lambda _3 + \sum_{j=1}^3 N_{3j} x_j^2 \right] \nonumber
\end{eqnarray}
Here the dependent variables $x_i$ describe the amplitudes of the three 
interacting modes; the $\lambda$'s, when negative, account for dissipation; 
the $N_{ij}$ denote modes of competition among waves; finally, the model 
includes a resonance term, which is that dependent on $\gamma$. 

Equations (\ref{s3ob}) have ten nonlinear terms. A good possibility for 
simplifying them proceeds without embedding, by means of a quasimonomial 
transformation of matrix:
\begin{equation}
   \label{s3oqm}
   C = \left( \begin{array}{ccc}
               N_{11} & N_{12} & N_{13}  \\
               N_{21} & N_{22} & N_{23}  \\
               N_{31} & N_{32} & N_{33}
              \end{array} \right) 
\end{equation}
The new vector field is given by matrices of the form:
\begin{equation}
   \label{s3ob3}
          M' = \left( \begin{array}{ccccc} 
               \lambda '_1 & 1 & 0 & 0 & \gamma '_1 \\ 
               \lambda '_2 & 0 & 1 & 0 & \gamma '_2 \\
               \lambda '_3 & 0 & 0 & 1 & \gamma '_3 
              \end{array} \right) \;\:\; , \;\;\:
          B' = \left( \begin{array}{ccc}
               2N_{11} & 2N_{12} & 2N_{13}  \\
               2N_{21} & 2N_{22} & 2N_{23}  \\
               2N_{31} & 2N_{32} & 2N_{33}  \\
               \nu _1  & \nu _2  & \nu _3
              \end{array} \right) \;\; ,
\end{equation}
where $\nu _j = -N_{1j}+N_{2j}+N_{3j}$, and the new system will be:
\begin{eqnarray}
   \dot{y}_1 & = & y_1 [ \lambda '_1 + 
         y_1^{2N_{11}}y_2^{2N_{12}}y_3^{2N_{13}} + 
         \gamma '_1 y_1^{\nu _1}y_2^{\nu _2}y_3^{\nu _3}] \nonumber \\
   \dot{y}_2 & = & y_2 [ \lambda '_2 + 
         y_1^{2N_{21}}y_2^{2N_{22}}y_3^{2N_{23}} +
         \gamma '_2 y_1^{\nu _1}y_2^{\nu _2}y_3^{\nu _3}] \label{s3ob4} \\
   \dot{y}_3 & = & y_3 [ \lambda '_3 + 
         y_1^{2N_{31}}y_2^{2N_{32}}y_3^{2N_{33}} +
         \gamma '_3 y_1^{\nu _1}y_2^{\nu _2}y_3^{\nu _3}] \nonumber
\end{eqnarray}
The new variables are $ y_i = x_1^{d_{i1}}x_2^{d_{i2}}x_3^{d_{i3}}$, where 
the $d_{ij}$ are the entries of $D=C^{-1}$. Therefore, they are nonlinear 
combinations of the original amplitudes $x_i$, the exponents being functions 
of the competition coefficients $N_{ij}$. Then, our 
redefinition of the amplitudes is such that for the new, equivalent set of 
amplitudes $y_i$, the competition with the other waves is summarized in only 
{\em one\/} nonlinear term per wave. This new representation is not optimal 
from the point of view of the resonance, which is now present in all 
equations. The overall result, however, is that of a simplification, since 
the number of nonlinear terms in the new flow is only six.

The remaining possibility is that of a complete embedding: 
\begin{equation}
   \tilde{M} = \left( \begin{array}{ccccc} 
               \lambda _1 & N_{11} & N_{12} & N_{13} & \gamma \\ 
               \lambda _2 & N_{21} & N_{22} & N_{23} &  0     \\
               \lambda _3 & N_{31} & N_{32} & N_{33} &  0     \\
                   0      &   0    &    0   &   0    &  1  
              \end{array} \right) \;\:\; , \;\;\:
   \tilde{B} = \left( \begin{array}{cccc}
              2 & 0 & 0 & 0 \\ 
              0 & 2 & 0 & 0 \\ 
              0 & 0 & 2 & 0 \\ 
             -1 & 1 & 1 & 0 
              \end{array} \right) 
\end{equation}
The quasimonomial transformation now is evidently that of matrix $C = 
\tilde{A}$. The resulting matrices are of the form:
\begin{equation}
  \tilde{M}' = \left( \begin{array}{ccccc} 
               \tilde{\lambda }_1 & 1 & 0 & 0 & 0 \\ 
               \tilde{\lambda }_2 & 0 & 1 & 0 & 0 \\
               \tilde{\lambda }_3 & 0 & 0 & 1 & 0 \\
               \tilde{\lambda }_4 & 0 & 0 & 0 & 1  
              \end{array} \right) \;\:\; , \;\;\:
  \tilde{B}' = \left( \begin{array}{cccc}
               2N_{11} & 2N_{12} & 2N_{13} & 2 \gamma \\
               2N_{21} & 2N_{22} & 2N_{23} & 0  \\
               2N_{31} & 2N_{32} & 2N_{33} & 0  \\
               \nu _1  & \nu _2  & \nu _3  & - \gamma 
              \end{array} \right) 
\end{equation}
The system has thus achieved the maximal simplification:
\begin{eqnarray}
   \dot{z}_1 & = & z_1 [ \tilde{ \lambda }_1 + 
         z_1^{2N_{11}}z_2^{2N_{12}}z_3^{2N_{13}}z_4^{2 \gamma}] \nonumber \\
   \dot{z}_2 & = & z_2 [ \tilde{ \lambda }_2 + 
         z_1^{2N_{21}}z_2^{2N_{22}}z_3^{2N_{23}}] \nonumber \\
   \dot{z}_3 & = & z_3 [ \tilde{ \lambda }_3 + 
         z_1^{2N_{31}}z_2^{2N_{32}}z_3^{2N_{33}}] \label{s2of} \\
   \dot{z}_4 & = & z_4 [ \tilde{ \lambda }_4 + 
         z_1^{\nu _1}z_2^{\nu _2}z_3^{\nu _3}z_4^{- \gamma}] \nonumber 
\end{eqnarray}
From the physical point of view, both types of interactions among waves 
(competition among modes and resonance) are now described simultaneously by 
one single nonlinear term in each of the equations. This can be clearly 
inferred from the form of the exponents of the quasimonomials in (\ref{s2of}). 
In this case, the simplification of the number of nonlinear terms is the 
maximal possible, since only four of them are present in system (\ref{s2of}). 

From the form of matrix $C$, we need not compute explicitly the 
projection matrix to find the projection plane: It must be $\{ \ln(z_4) = 
0 \}$, since the plane $\{ \ln(x_4) = 0 \}$ is invariant under the 
quasimonomial transformation. 

\begin{flushleft}
{\bf 5. Final remarks}
\end{flushleft}

We may conclude by a few remarks on the unified scheme presented here when 
compared to other approaches to the case problems. We mentioned in the 
Introduction the work of Kerner \cite{ker} on the reduction of general 
systems of ODEs to what he calls ``Elemental Riccati systems'' (ERS), the 
vector field of which is strictly quadratic ---with simple arrays of 
coefficients that are either zero or unity. Let us consider Kerner's 
example, which starts from the one-dimensional standard cubic equation:
\begin{equation}
   \dot{x} = a_0 + a_1 x + a_2 x^2 + a_3 x^3 \;\:\;\: , 
   \label{kex}
\end{equation}
and produces an eleven-dimensional ERS
\begin{equation}
   \dot{z}_{ \alpha } = \sum_{ \beta, \gamma = 1}^{11} E^{*}_{ \alpha \beta 
   \gamma } z_{ \beta } z_{ \gamma } \; , \;\:\; \alpha = 1 , \ldots , 11
\end{equation}
The QP formalism, when applied to (\ref{kex}), leads to a 
three-dimensional Lotka-Volterra system, a much smaller dimensionality, 
indeed. Both the ERS and the Lotka-Volterra systems are compact 
formats, liable to be analyzed in a purely algebraic context.
Notwithstanding these similarities, there is a particularity 
which favors the QP formalism. Kerner's approach proceeds in two steps: A 
rule of thumb transformation to a Riccati system, followed by a sequence of 
algebraic operations to produce the final ERS form. On the contrary, in the 
QP formalism the procedure is systematically carried out from the very 
beginning through simple matrix operations.

Finally, a similar assessment can be made when it comes to comparing known 
alternative approaches for a simplification of the number of nonlinear terms 
in each component of the flow. We may recall the {\em ad hoc\/} method of 
Savageau and collaborators \cite{vsi,vo1}, a heuristic technique which raises 
serious doubts about its mathematical consistency. It strongly relies on 
arbitrary user-dependent choices of additional variables that lead to 
manifold systems complying to a specified format. This is in opposition to 
the unique unimonomial form we are able to construct by the technique of 
Subsection 4.1.

\mbox{}

\mbox{}

{\em Acknowledgements: 
This work has been supported by the DGICYT (Spain), under grant PB94-0390, 
and by the EU Esprit WG 24490 (CATHODE-2). B. H. acknowledges a doctoral 
fellowship from Comunidad Aut\'{o}noma de Madrid. The authors also 
acknowledge valuable suggestions from anonymous referees.} 

\pagebreak

\end{document}